\documentclass[twocolumn]{aastex61}

\newcommand{\lsim}{\raisebox{-0.13cm}{~\shortstack{$<$ \\[-0.07cm] $\sim$}}~}
\newcommand{\gsim}{\raisebox{-0.13cm}{~\shortstack{$>$ \\[-0.07cm] $\sim$}}~}

\received{...}
\revised{...}
\accepted{...}
\submitjournal{ApJ}

\shorttitle{A clear starburst/main-sequence bimodality for $\rm H\alpha$ emitters at $\lowercase{z}\sim4-5$}
\shortauthors{Caputi et al.}
\begin{document}

\title{Star formation in galaxies at $\lowercase{z}\sim4-5$ from the SMUVS survey:  a clear starburst/main-sequence bimodality for $\rm H\alpha$ emitters on the $\rm SFR-M^\ast$ plane}

\correspondingauthor{K. I. Caputi}
\email{karina@astro.rug.nl}

\author{K. I. Caputi}
\affil{Kapteyn Astronomical Institute, University of Groningen, P.O. Box 800, 9700AV Groningen, The Netherlands}

\author{S. Deshmukh}
\affil{Kapteyn Astronomical Institute, University of Groningen, P.O. Box 800, 9700AV Groningen, The Netherlands}

\author{M. L. N. Ashby}
\affil{Harvard-Smithsonian Center for Astrophysics, 60 Garden St., Cambridge, MA 02138, USA}

\author{W. I. Cowley}
\affil{Kapteyn Astronomical Institute, University of Groningen, P.O. Box 800, 9700AV Groningen, The Netherlands}

\author{L. Bisigello}
\affil{Kapteyn Astronomical Institute, University of Groningen, P.O. Box 800, 9700AV Groningen, The Netherlands}

\author{G. G. Fazio}
\affil{Harvard-Smithsonian Center for Astrophysics, 60 Garden St., Cambridge, MA 02138, USA}

\author{J. P. U. Fynbo}
\affil{Dark Cosmology Centre, Niels Bohr Institute, University of Copenhagen, Juliane Maries Vej 30, 2100 Copenhagen, Denmark}

\author{O. Le F\`evre}
\affil{Aix Marseille Universit\'e, CNRS, LAM (Laboratoire d'Astrophysique de Marseille), UMR 7326, 13388, Marseille, France}

\author{B. Milvang-Jensen}
\affil{Dark Cosmology Centre, Niels Bohr Institute, University of Copenhagen, Juliane Maries Vej 30, 2100 Copenhagen, Denmark}

\author{O. Ilbert}
\affil{Aix Marseille Universit\'e, CNRS, LAM (Laboratoire d'Astrophysique de Marseille), UMR 7326, 13388, Marseille, France}

%% Note that the \and command from previous versions of AASTeX is now
%% depreciated in this version as it is no longer necessary. AASTeX 
%% automatically takes care of all commas and "and"s between authors names.

%% AASTeX 6.1 has the new \collaboration and \nocollaboration commands to
%% provide the collaboration status of a group of authors. These commands 
%% can be used either before or after the list of corresponding authors. The
%% argument for \collaboration is the collaboration identifier. Authors are
%% encouraged to surround collaboration identifiers with ()s. The 
%% \nocollaboration command takes no argument and exists to indicate that
%% the nearby authors are not part of surrounding collaborations.

%% Mark off the abstract in the ``abstract'' environment. 
\begin{abstract}

We study a large galaxy sample from the {\em Spitzer Matching Survey of the UltraVISTA ultra-deep Stripes (SMUVS)}  to search for sources with enhanced $3.6 \, \rm \mu m$ fluxes indicative of strong $\rm H\alpha$ emission at $z=3.9-4.9$.  We find that the percentage of ``$\rm H\alpha$ excess'' sources reaches  37-40\% for galaxies with stellar masses $\rm log_{10} (M^\ast \rm / M_\odot) \approx 9-10$, and decreases to $<20\%$ at  $\rm log_{10} (M^\ast \rm / M_\odot) \sim 10.7$. At higher stellar masses, however, the trend reverses, although this is likely due to AGN contamination. We derive star formation rates (SFR) and specific SFR (sSFR) from the inferred $\rm H\alpha$ equivalent widths (EW) of our ``$\rm H\alpha$ excess'' galaxies. We show, for the first time, that the  ``$\rm H\alpha$ excess'' galaxies clearly have a bimodal distribution on the SFR-$\rm M^\ast$ plane: they lie on the main sequence of star formation (with $\rm log_{10} (sSFR \rm / yr^{-1})<-8.05$) or in a starburst cloud (with $\rm log_{10} (sSFR \rm / yr^{-1}) >-7.60$). The latter contains $\sim 15\%$ of all the objects in our sample and accounts for $>50\%$ of the cosmic SFR density at $z=3.9-4.9$, for which we derive a robust lower limit of $0.066 \, \rm M_\odot \, yr^{-1} \, Mpc^{-3}$. Finally, we identify an unusual $>50\sigma$ overdensity of $z=3.9-4.9$ galaxies within a $0.20 \times 0.20$~arcmin$^2$ region. We conclude that the SMUVS unique combination of area and depth at mid-IR wavelengths provides an unprecedented level of statistics and dynamic range which are fundamental to reveal new aspects of galaxy evolution in the young Universe.

 \end{abstract}

%% Keywords should appear after the \end{abstract} command. 
%% See the online documentation for the full list of available subject
%% keywords and the rules for their use.

%% From the front matter, we move on to the body of the paper.
%% Sections are demarcated by \section and \subsection, respectively.
%% Observe the use of the LaTeX \label
%% command after the \subsection to give a symbolic KEY to the
%% subsection for cross-referencing in a \ref command.
%% You can use LaTeX's \ref and \label commands to keep track of
%% cross-references to sections, equations, tables, and figures.
%% That way, if you change the order of any elements, LaTeX will
%% automatically renumber them.

%% We recommend that authors also use the natbib \citep
%% and \citet commands to identify citations.  The citations are
%% tied to the reference list via symbolic KEYs. The KEY corresponds
%% to the KEY in the \bibitem in the reference list below. 

\section{Introduction} \label{sec:intro}

Studying star formation in galaxies at high redshifts is crucial to understanding the early stages of galaxy evolution. Over the last ten years, a picture has emerged indicating that the global cosmic star formation rate (SFR) density increased after the Big Bang until reaching a peak about ten billion years ago, and then declined until today \citep[e.g.][]{hop06,beh13,mad14}. That peak is likely the product of two effects: mainly the net increase in the number density of galaxies that make the bulk of star formation in the first few billion years and possibly, but this is less clear, the fact that the SFR of individual galaxies may increase over that time.  Indeed,  our current knowledge of {\em how} star formation and stellar mass buildup proceeded over the first few billion years is still very sparse. 

In most observed star-forming galaxies up to at least $z\sim3$, the instantaneous SFR appears to correlate (within some scatter) with the stellar mass.  This correlation on the SFR-$\rm M^\ast$ plane  is the so-called `galaxy star-formation main sequence' \citep[e.g.,][]{noe07,rod10,elb11}. For these galaxies the specific star formation rates (sSFR$\equiv \rm SFR/\rm M^\ast$) are roughly constant, which implies the existence of a scaling relation between gas consumption and galaxy growth \citep[see e.g.,][]{pop15,lag16}. In addition, there exists a minority of star-forming galaxies which are characterised by significantly higher sSFRs and are usually called `starbursts'. The starburst phase is presumably a temporary state in which the galaxy is taken out of the main sequence, due to some kind of perturbation that temporarily enhances the star formation. Starburst galaxies are very rare in the local Universe and have been found to constitute a small fraction of the dusty star-forming galaxies observed at redshifts $z\sim0.3$ to $z\sim2$ \citep[][]{rod11,sar12}.   This inferred weak evolution in the starburst fraction with redshift is based on the analysis of the most luminous dusty galaxies, as only these galaxies could be detected by the last generation of infrared (IR) telescopes, particularly the {\em Herschel Space Telescope}. Therefore, these results mainly concern massive galaxies and it is not obvious whether they can be extrapolated to all star-forming galaxies.

A main limiting factor to understanding galaxy evolution in the high-$z$ Universe has been the lack of deep galaxy surveys over significantly large areas of the sky. Such surveys could provide sufficient statistics and dynamic range to investigate how star formation, as well as different physical properties, vary among different stellar-mass galaxies. For star formation, in particular, another limitation is that the only tracer currently available for large galaxy samples is the rest-ultraviolet (UV) flux shifted into observed optical wavelengths at high-$z$. UV fluxes are extremely sensitive to dust extinction and, thus, the derived SFRs can be very uncertain. Far-IR and (sub)-millimetre photometry are the ideal complement to UV photometry for the purpose of obtaining total SFR estimates, but  single-dish far-IR telescopes are insufficiently sensitive to systematically study representative galaxy samples in the high-$z$ Universe. As an alternative, Balmer lines, especially $\rm H\alpha \, \lambda 6563$, which is much less affected by dust than the UV spectral continuum, is a very suitable SFR tracer widely used for galaxies up to $z\sim2-3$. However, $\rm H\alpha$ is shifted into the mid-IR regime ($\lambda \gsim 3 \, \rm \mu m$) at $z>3$, making its detection prohibitive for current spectrographs.

When emission lines have sufficiently large equivalent widths (EW), however, their presence can be inferred even from broad-band photometry. At redshifts $3.9 \lsim z \lsim 4.9$, the $\rm H\alpha$ line is encompassed by the $3.6 \, \rm \mu m$ filter passband at the Infrared Array Camera \citep[IRAC; ][]{faz04} on board the {\em Spitzer Space Telescope} \citep{wer04}. This fact has been exploited by different authors to investigate the presence of intense $\rm H\alpha$ emitters at these redshifts. \citet{shi11} analyzed a sample of 74 galaxies at similar redshifts and concluded that $\sim 70\%$ of these galaxies have an excess flux at $3.6  \, \rm \mu m$ with respect to the stellar continuum. More recently, \citet{sta13} and \citet{smi16} conducted similar analyses, based on galaxy samples with spectroscopic and photometric redshifts, and confirmed the presence of intense $\rm H\alpha$ emitters at $z\sim4-5$. The percentage of galaxies displaying an ``$\rm H\alpha$ excess'' and the derived rest EWs varied according to the different selection effects, but in general they found median values of $\rm H\alpha$ EW$\approx300-400 \, \rm \AA$. These EWs are on average much larger than those observed in the local Universe and are broadly consistent with an increase with redshift, as determined by \citet{fum12}, \citet{fai16} and \citet{mar16}. All these studies have been very valuable to raise awareness on the increasing importance of nebular emission up to at least $z\sim5$. However, none of them has analyzed a sufficiently representative galaxy sample at $z \sim 4-5$, which would allow for a more complete investigation of the implications of these results within the context of galaxy evolution.

The {\em Spitzer Matching Survey of the Ultra-VISTA ultra-deep Stripes} (SMUVS; M.~Ashby et al., 2017, in preparation) is a {\em Spitzer} Exploration Science Program which has collected ultra-deep IRAC $3.6$ and $4.5 \, \rm \mu m$ images over a significant part of the COSMOS field \citep{sco07}, making it the largest quasi-contiguous {\em Spitzer} field to analyze the high-$z$ Universe.  The SMUVS unprecedented level of statistics allows for a detailed study of galaxy properties and evolution over more than three decades in stellar mass at high $z$. In this paper we analyze the large SMUVS galaxy sample containing almost 6000 sources at $3.9 \leq z \leq 4.9$ to investigate the presence of prominent $\rm H\alpha$ emitters as a function of stellar mass, along with the implications for the SFR versus $\rm M^\ast$ relation at those high redshifts. 

This paper is organised as follows. In Section \S\ref{sec:data} we describe the utilised datasets, source catalogue construction, as well as photometric redshifts and stellar mass derivations. In Section \S\ref{sec:samplesel} we explain our selection criteria for  ``$\rm H\alpha$ excess'' galaxies among all SMUVS galaxies at $3.9 \leq z \leq 4.9$. In Section \S\ref{sec:results} we present our results on the derived nebular line equivalent widths;  SFRs and resulting SFR-$\rm M\ast$ and sSFR-$\rm M\ast$ relations; and the inferred cosmic star formation rate density at $z\sim4-5$. We also discuss a rare $z\sim4-5$ overdensity in the SMUVS field. Finally, in Section \ref{sec:concl} we summarize our findings and discuss our conclusions. Throughout this paper we adopt a cosmology  with $\rm H_0=70 \,{\rm km \, s^{-1} Mpc^{-1}}$, $\rm \Omega_M=0.3$ and $\rm \Omega_\Lambda=0.7$. All magnitudes and fluxes are total, with magnitudes referring to the AB system \citep{oke83}. Stellar masses correspond to a \citet{cha03} initial mass function (IMF), except where explicitly stated otherwise (see Section \S\ref{subsec:sfrd}).

\section{Datasets and Photometric Redshifts} \label{sec:data}

The SMUVS program (PI Caputi; M.~Ashby et al., 2017, in preparation) has collected ultra-deep {\em Spitzer} 3.6 and 4.5~$\rm \mu m$ data over the region of the COSMOS field \citep{sco07} overlapping the three UltraVISTA ultra-deep stripes \citep{mcc12} with deepest optical coverage from the Subaru telescope \citep{tan07}. The SMUVS mosaics utilised here correspond to the survey almost final depth, which reaches an average integration time of $\sim 25 \, \rm h$/pointing \citep[including previously available IRAC data in COSMOS;][]{san07,ash13,ste14,ash15}. The considered UltraVISTA data correspond to the third data release (DR3), which in the ultra-deep stripes reaches an average depth of $K_s=24.9 \pm 0.1$ and $H=25.1\pm 0.1$  ($2^{\prime\prime}$ diameter; $5\sigma$)\footnote{see http://www.eso.org/sci/observing/phase3/data\_releases/ \\ uvista\_dr3.pdf}.

A complete description of our SMUVS source multi-wavelength catalogue construction and spectral energy distribution (SED) fitting is provided in S.~Deshmukh et al.~(2017, in preparation). Here we only summarize our main steps. Firstly, we extracted sources on UltraVISTA $HKs$ average stack mosaics of the three relevant ultra-deep stripes, using the software SExtractor \citep{ber96} with a detection threshold of 1.5$\sigma$ over 5 contiguous pixels. The position of these sources were considered as priors to perform iterative PSF-fitting photometric measurements on the {\em Spitzer} SMUVS 3.6 and 4.5~$\rm \mu m$ mosaics, using the DAOPHOT package on IRAF with empirical point-spread-functions (PSFs) obtained from stars in the field (in each stripe separately).  The PSF-fitting algorithm converged for $\sim70\%$ of the sources. For the remaining ones,  we measured directly IRAC aperture fluxes in $2.4 \, \rm arcsec$-diameter circular apertures at the position of the UltraVISTA sources, and corrected them to total fluxes multiplying by a factor of 2.13, which was determined from the curves of growth of stars in the field. Overall, we found that $\sim 95-96\%$ of the UltraVISTA ultra-deep sources are detected in at least one IRAC band (and $93-94\%$ in both bands).  The comparison of the resulting IRAC number counts with those obtained in the deeper S-CANDELS survey \citep{ash15} indicates that our resulting SMUVS catalogue is 80\%(50\%)  complete at [3.6] and [4.5]=25.5 (26.0) total magnitudes. 

For all these sources, we measured 2-arcsec diameter circular photometry on 26 broad, intermediate and narrow bands from the $U$ through the $K_s$ bands, using SExtractor on dual-image mode with the UltraVISTA $HKs$ stacks as detection images, and applied corresponding point-source aperture corrections in each band. After cleaning for galactic stars based on a ($J$-[3.6]) versus ($B$-$J$) colour-colour diagram \citep[e.g.,][]{cap11} and masking regions of contaminated light around the brightest sources, our final catalogue contains $\sim 291,300$ UltraVISTA sources with at least one IRAC-band detection over a net area of 0.66~deg$^2$. This is our SMUVS parent catalogue with 28-band photometry ($U$ through $K_s$ + IRAC) for SED fitting analysis.

The PSF-fitting technique assumes that all sources are point-like: on the IRAC images, this is indeed a reasonable assumption for virtually all sources with $[3.6]>21$~mag \citep[see Fig.~25 in][]{ash13}. Besides, all the multi-wavelength photometry that we considered for SED fitting has been measured on circular apertures (and corrected to total), so the IRAC photometry based on PSF fitting is consistent with this procedure. 

We performed the source SED fitting using the $\chi^2$-minimization code LePhare \citep{arn99,ilb06} over our 2-arcsec-based total flux source catalogue. We used Bruzual \& Charlot~(2003) templates corresponding to a single stellar population and different exponentially declining star formation histories (SFHs), all with solar metallicity ($\rm Z_\odot$), allowing for the addition of emission lines. In Appendix A, we discuss the impact of considering two possible metallicities ($\rm Z_\odot$ and $\rm 0.2 \, Z_\odot$) for the SED fitting.  To account for internal dust extinction, we used the \citet{cal00} reddening law.  Throughout this paper we will consider that the extinction derived in this way from the SED fitting of each galaxy is the same for the spectral continuum and lines at the same wavelengths, which should be a reasonable assumption for the majority of our galaxies \citep[e.g.,][]{red10,shi15}. In Appendix B, we explore the effects of assuming a different dust-extinction law which is directly dependent on the UV slope of each galaxy. 

As in \citet{cap15}, in the case of non-detections we adopted $3\sigma$ flux upper limits in the broad bands and ignored narrow and intermediate bands. For the minority ($<14\%$) of sources corresponding to [3.6] or [4.5]$>22$ galaxies with a [3.6] or [4.5]$<23$ neighbour within a 3~arcsec radius, we performed the SED fitting only on 26 bands (ignoring the IRAC bands).

We obtained zero-point corrections to the photometry from a first LePhare run and then we performed a second, definitive run taking into account these corrections. From LePhare's output we obtained photometric redshifts and stellar mass estimates for $> 99.9\%$ of our sources. The COSMOS field has a large amount of spectroscopic data which are very useful to assess the quality of the obtained photometric redshifts \citep[e.g.,][]{lil07}: the resulting dispersion of the $|z_{\rm phot}$-$z_{\rm spec}|$$/(1+z_{\rm spec})$ distribution in our sample, based on $\sim$14,000 galaxies with reliable spectroscopic redshifts, is $\sigma=0.026$ with 5.5\% outliers (S. Deshmukh et al., 2017, in preparation). This general photometric redshift quality is similar to that restricted to the redshift range analyzed here, i.e., $3.9 \leq z \leq 4.9$. At these high redshifts we have 55 sources with spectroscopic redshifts and obtain $\sigma=0.023$ with 7.3\% outliers.

From the obtained SMUVS photometric redshift catalogue, we excluded $\sim 1\%$ of sources because their best-fit $z_{\rm phot}$ were incompatible with their detection at short wavelengths -- see criteria in \citet{cap15}. Our final SMUVS catalogue with photometric redshift determinations contains 288,003 galaxies, including 5925 at $3.9 \leq z\leq 4.9$, which is the relevant redshift range in this work. As part of LePhare's output, we also have stellar mass estimates for all these sources.

\section{Selection of prominent $\rm H\alpha$ emitters at $\lowercase{z}=3.9-4.9$} \label{sec:samplesel}

At redshifts $3.9 \leq z \leq 4.8$  the entire ($\rm H\alpha$ $\lambda6563$+ \\ +[NII]~$\lambda\lambda6548,6583$+ [SII]~$\lambda\lambda6716, 6730$)  emission line complex is encompassed within the IRAC $3.6 \, \rm \mu m$ filter and the ($\rm H\alpha$~$\lambda6563$ + [NII]~$\lambda\lambda6548, 6583$) complex is present up to $z \approx 4.9$.  At these same redshifts, no prominent emission line is present in the IRAC $4.5 \, \rm \mu m$ filter. At $z\gsim4.4$, the [OII]~$\lambda\lambda3727$ enters the $K_s$  band, although its EW is typically significantly smaller than the $\rm H\alpha$ EW \citep[e.g.,][]{mou05}. The ($\rm H\beta$~$\lambda 4861$ + [OIII]~$\lambda\lambda 5007$) complex lies in the gap between the $K_s$  and $3.6 \, \rm \mu m$ bands. 

Only emission lines or line complexes with sufficiently large EW can produce a significant flux excess in a broad-band. For the SED model templates considered in this work, we determined that producing a $3.6 \, \rm \mu m$ magnitude brightening of at least $0.1$~mag with respect to the continuum at $3.9 \lsim z \lsim 4.9$ requires a ($\rm H\alpha$ + [NII] + [SII]) rest-frame $\rm EW \gsim 150 \, \AA$. 

With all these considerations in mind,  in order to determine which of the 5925 galaxies at $3.9 \leq z \leq 4.9$  have a significant $3.6 \, \rm \mu m$  photometric excess with respect to the continuum, we re-modelled their SEDs running LePhare again {\em without} emission lines,  excluding the $3.6 \, \rm \mu m$ band, and adopting the fixed ${z_{\rm phot}}$ that were previously determined in the original run (with all bands and emission lines).  We selected the ``$\rm H\alpha$ excess'' sources by imposing that: 
\vspace{0.3cm}

\noindent 1) $\rm [3.6]^{obs}-[3.6]^{mod} < -0.1$, where $\rm [3.6]^{obs}$ is the observed $3.6 \, \rm \mu m$ magnitude and $\rm [3.6]^{mod}$ is the filter-convolved model magnitude obtained from the new LePhare SED fitting without emission lines; 

\vspace{0.2cm}
\noindent 2) $\rm [4.5]^{obs}-[4.5]^{mod} > -0.1$, so there is no excess light at $4.5 \, \rm \mu m$; 

\vspace{0.2cm}
\noindent 3) $K_s^{\rm obs}$-$K_s^{\rm mod} > -0.1$, so there is no excess light in the $K_s$ band either.

\vspace{0.2cm}
\noindent 4) if the source has any IRAC magnitude $>22$, then it should not have any neighbour with an IRAC magnitude $<23$ within a 3~arcsec radius.

\vspace{0.3cm}
These conditions altogether make for a conservative approach that guarantees that the continuum around the ($\rm H\alpha$ + [NII] + [SII]) complex is well modelled and we do not overestimate the $3.6 \, \rm \mu m$-band flux excess. Although we did not set upper limits for the  $K_s^{\rm obs}$-$K_s^{\rm mod}$ and $\rm [4.5]^{obs}-[4.5]^{mod}$ differences, we note that their median values are $\sim 0.12$ and $\sim -0.005$~mag, respectively, i.e., comparable or lower than the typical photometric error bars, i.e., the observed fluxes are consistent with the continuum level. 

We found that 1904 galaxies satisfy conditions 1) to 4), i.e., about a third of our SMUVS galaxy sample at $3.9 \leq z \leq 4.9$. In the rest of this paper, we refer to these galaxies as the ``$\rm H\alpha$ excess'' sample.

\section{Results} \label{sec:results}

\subsection{($H\alpha$ + [NII] + [SII]) emission versus stellar mass in galaxies at $z=3.9-4.9$\label{subsec:haew}}

\begin{figure*}[ht!]
\center{
\includegraphics[width=1\linewidth, keepaspectratio]{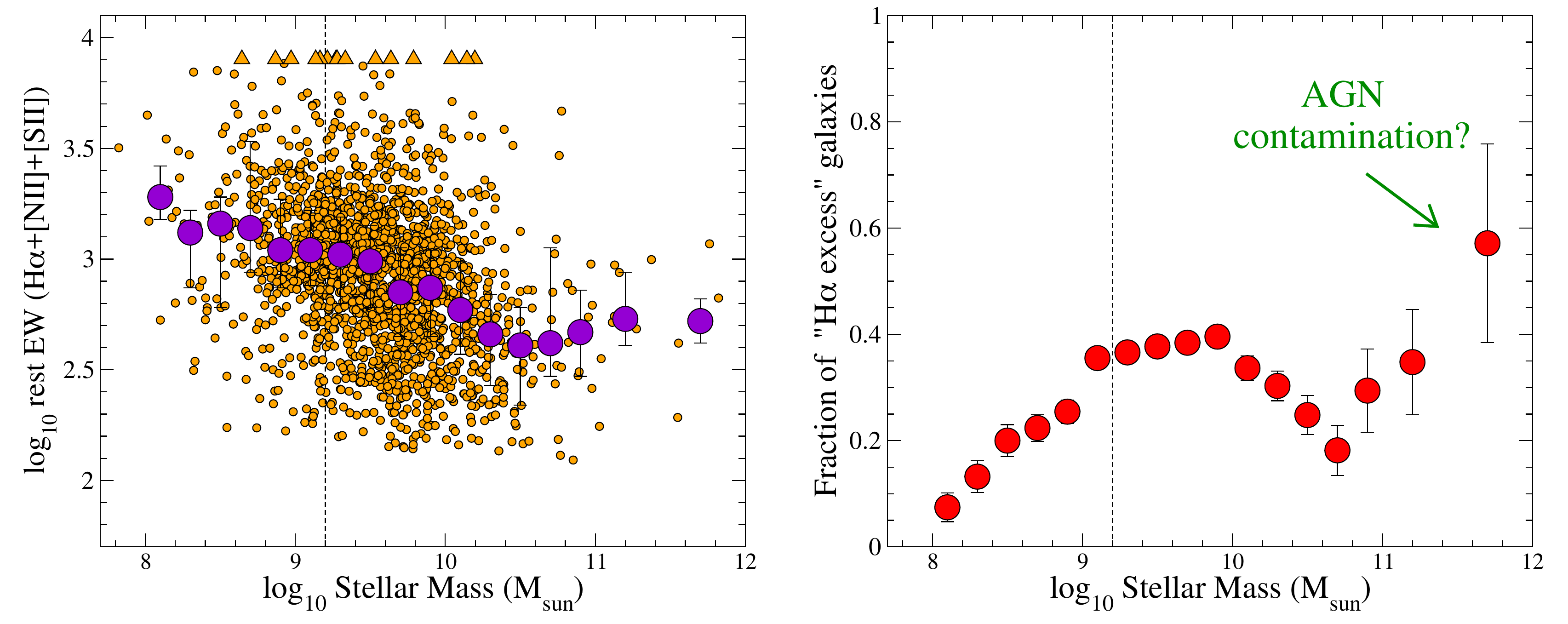}
\caption{{\em Left:} ($\rm H\alpha$ + [NII] + [SII]) (($\rm H\alpha$ + [NII]) at $z>4.8$) rest EW versus stellar mass for the ``$\rm H\alpha$ excess''  galaxies at $3.9 \leq z \leq 4.9$. The big purple circles show the median EW values at different stellar masses and the error bars indicate the two central quartiles of the EW distribution in each stellar mass bin. The vertical dashed line indicates the SMUVS 50\%  stellar-mass completeness at  $3.9 \leq z \leq 4.9$. {\em Right:} Fraction of SMUVS $3.9 \leq z \leq 4.9$ galaxies with  ``$\rm H\alpha$ excess'' versus stellar mass. }. \label{fig_ewvsstm}
}
\end{figure*}

We converted the $3.6 \, \rm \mu m$  flux excess of each galaxy with ``$\rm H\alpha$ excess'' into the corresponding ($\rm H\alpha$ + [NII] + [SII]) rest-frame EW. For galaxies at $z>4.8$, the excess corresponds to ($\rm H\alpha$ + [NII]) only.  We constructed a grid of $3.6 \, \rm \mu m$  photometric excesses corresponding to different rest $\rm EW$  for each galaxy SED template (characterised by a SFH and age) at different redshifts. To do this, we modelled each line complex as a single Gaussian, with rest-frame widths of $40$ and $180 \, \rm \AA$ for the ($\rm H\alpha$ + [NII]) and ($\rm H\alpha$ + [NII] + [SII]), respectively. Note that the exact values of these widths are irrelevant, as they cancel out when recovering the corresponding line EW and fluxes, as discussed by \citet{shi11}. We convolved the galaxy SED templates, each with an added line complex of different rest EW,  with the IRAC $3.6 \, \rm \mu m$ filter transmission curve. In each case, we measured the resulting IRAC $3.6 \, \rm \mu m$ magnitude and  worked out the magnitude difference with respect to the original SED template with no emission lines. Finally, we considered each real galaxy photometric redshift and best SED model parameters to infer the corresponding line complex  $\rm EW$ from the observed $3.6 \, \rm \mu m$  flux excess, by interpolating the values in the model grid.

Figure~\ref{fig_ewvsstm} (left) shows the derived ($\rm H\alpha$ + [NII] + [SII]) (($\rm H\alpha$ + [NII]) at $z>4.8$)  rest EW versus stellar mass for our ``$\rm H\alpha$ excess''  galaxies. The stellar masses shown in this and all the remaining plots in this paper are those obtained from the original LePhare SED fitting (i.e., those including the $3.6 \, \rm \mu m$ band and line emission). We see that these rest $\rm EW$ span values between $\sim 150 \, \rm \AA$ and several thousand $\rm \AA$. The median values vary with stellar mass: they are $\rm \langle EW \rangle \approx1000 \, \AA$ for galaxies with $\rm M^\ast=(1-4) \times 10^9 \, M_\odot$, but only $\rm \langle EW \rangle\approx400 \, \AA$ for galaxies with $\rm M^\ast=(1.5-4) \times 10^{10} \, M_\odot$. These values are broadely consistent with those obtained by \citet{smi16}.

The incidence of ``$\rm H\alpha$ excess'' sources also changes with stellar mass. As we can see from Figure~\ref{fig_ewvsstm} (right), the fraction of sources with  ``$\rm H\alpha$ excess'' reaches $37\%-40\%$ at $\rm M^\ast=10^9 - 10^{10}\, M_\odot$, but decreases to $\approx 18\%$ at $\rm M^\ast\sim 5 \times 10^{10} \, M_\odot$.  As we discuss in next section, the fact that the percentage of galaxies with large $\rm H\alpha$ EW is higher among intermediate-mass galaxies indicates that the on-going with respect to past star formation is more important in them than in more massive galaxies at $3.9 \leq z \leq 4.9$.  

Interestingly, the ``$\rm H\alpha$ excess'' incidence rises again at $\rm M^\ast\gsim 6 \times 10^{10} \, M_\odot$, albeit with moderate median EW. This reversing trend is puzzling and may suggest that star formation does not subside in the most massive galaxies before it does in the more typical massive ones. As we discuss below, an alternative and more likely explanation is that the $3.6 \, \rm \mu m$ flux excess in the most massive galaxies is linked to the presence of active galactic nuclei (AGN).

\subsection{The SFR-$M^\ast$ and sSFR-$M^\ast$ planes\label{subsec:sfrstm}}

\subsubsection{SFR from $H\alpha$ emission}

\begin{figure*}[ht!]
\center{
\includegraphics[width=1\linewidth, keepaspectratio]{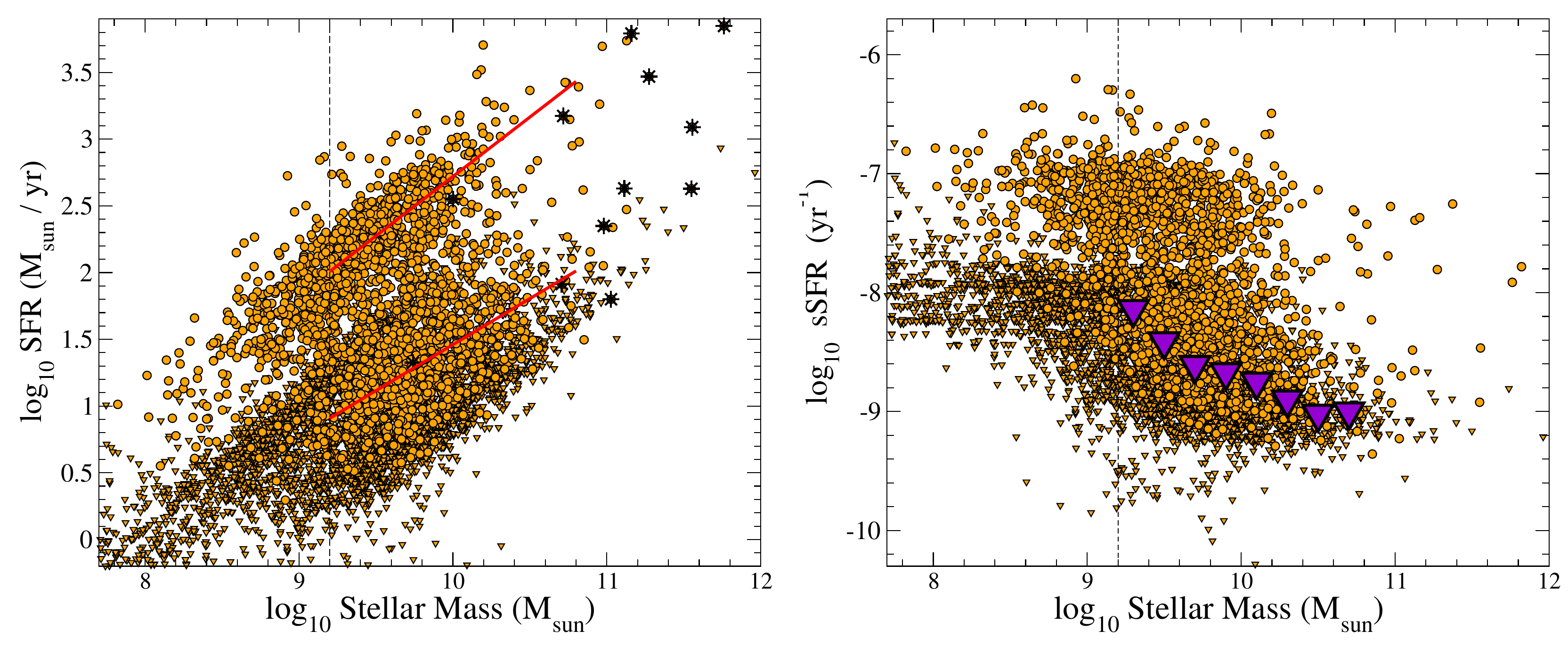}
\caption{{\em Left:} SFR based on the derived $\rm H \alpha$ luminosities for ``$\rm H\alpha$ excess'' galaxies (circles) and upper limits derived for galaxies with no $3.6 \, \rm \mu m$ excess (triangles), versus stellar mass, at $3.9 \leq z \leq 4.9$. The asterisks indicate galaxies that are $24 \, \rm \mu m$ detected, so they may have an AGN component in the IRAC bands. The ``$\rm H\alpha$ excess'' galaxies are clearly distributed in two clouds, which correspond to the galaxy main sequence and starbursts at those redshifts. The red lines show the results of linear regressions done on the two  clouds separately, considering only galaxies with $9.2 \leq \rm log_{10} (M^\ast) \leq 10.8$.  {\em Right:} the corresponding sSFR versus stellar mass diagram. The purple triangles indicate upper limits for the median sSFR per stellar mass bin, obtained considering {\em all} galaxies, i.e. the derived sSFR for ``$\rm H\alpha$ excess''  galaxies and upper limits for those with no  $3.6 \, \rm \mu m$ excess.} \label{fig_sfrvsstm}
}
\end{figure*}

From the ($\rm H\alpha$ + [NII] + [SII]) EW we can derive an $\rm H\alpha$-based star formation rate for each ``$\rm H\alpha$ excess'' galaxy. To do this, we obtained emission line fluxes assuming that the modelled SED continuum is constant over the entire line complex wavelength range. As explained in Section~\ref{sec:data}, we also assumed that the extinction is the same for the continuum and the lines at the same wavelengths, which should be reasonable for the bulk of our galaxies \citep[e.g.,][]{red10,shi15}. This extinction has been obtained from the SED fitting of each galaxy. To obtain the net $\rm H\alpha$ contribution to these fluxes, we considered that $f(\rm H\alpha)$$= 0.63 f(\rm H\alpha + [NII]+[SII])$ and $f(\rm H\alpha)$$= 0.81 f(\rm H\alpha + [NII])$, which is valid for solar metallicities  \citep{and03}. Then we converted the clean, derived $\rm H \alpha$ luminosities into SFRs using the \citet{ken98} relation:

\begin{equation}
\rm SFR (H\alpha) \, (M_\odot \, yr^{-1}) = 7.936 \times 10^{-42} \times L_{H\alpha} \, (erg \, s^{-1})
\label{eq-kenn}
\end{equation}

\noindent where the resulting SFR correspond to a \citet{sal55} IMF over (0.1-100)~$\rm M_\odot$, so we divided them by a factor 1.69 to re-scale them to a Chabrier IMF.

Figure~\ref{fig_sfrvsstm} shows the resulting SFR and sSFR versus stellar mass diagrams, which span more than three decades in stellar mass.  In these plots we include the SFR and sSFR based on the derived $\rm H \alpha$ luminosities for the ``$\rm H\alpha$ excess'' galaxies and upper limits for all other galaxies at $3.9 \leq z \leq 4.9$. These upper limits are based on the minimum line-complex EW that we can detect from flux excess in the $3.6 \, \rm \mu m$ band. The vertical dashed lines  indicate the 50\% stellar-mass completeness level of our sample, which is $\rm log_{10} M^\ast \approx 9.2$ at $3.9 \leq z \leq 4.9$. Note that the 80\% stellar-mass completeness level is $\rm log_{10} M^\ast\approx9.6$, but no main conclusion in this paper would change if we restricted our analysis to this higher stellar mass limit.

The asterisks on the SFR-$\rm M^\ast$ plot indicate galaxies that have a $24 \, \rm \mu m$ counterpart within a 2~arcsec radius in the Mid Infrared Photometer for {\em Spitzer} \citep[MIPS; ][]{rie04} S-COSMOS catalogue \citep{san07}. These sources correspond mostly to galaxies with $\rm log_{10} (M^\ast)> 10.8$. Given the limited depth of the $24 \, \rm \mu m$  catalogue, the detection of sources at $3.9 \leq z \leq 4.9$ suggests that they are likely AGN.  Unfortunately, an SED power-law analysis in the IRAC bands is not useful to study this issue further because of a k-correction effect: the maximum contribution of an AGN mid-IR power law is beyond the IRAC bands at such high redshifts, so only the hottest dust AGN could be manifested as IR power-law sources in  IRAC \citep{cap13}. But one of the $24 \, \rm \mu m$-detected most massive galaxies in our sample is indeed an X-ray source \citep{civ16} spectroscopically confirmed to be at $z=4.596$, which indicates its AGN nature. Therefore, as a matter of precaution, we flagged the $24 \, \rm \mu m$ detected sources in our sample and excluded all galaxies with $\rm log_{10} (M^\ast)> 10.8$ from further analysis. This high fraction of $24 \, \rm \mu m$ detections among the most massive  ``$\rm H\alpha$ excess'' galaxies supports our AGN hypothesis also based on the reversing trend discussed in Section~\ref{subsec:haew}, which shows that the $3.6 \, \rm \mu m$ incidence becomes higher at the highest mass end, after a continuous decrease from intermediate to high stellar masses.

\subsubsection{The bimodal distribution of $H\alpha$ emitters on the SFR-$M^\ast$ plane}

\begin{figure}[]
\center{
\includegraphics[width=1.1\linewidth, keepaspectratio]{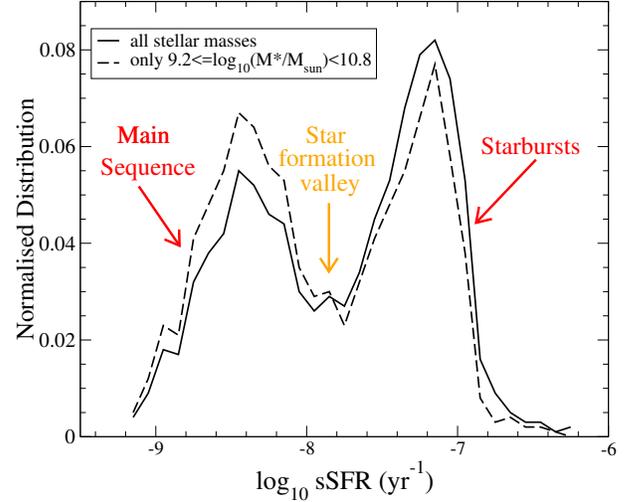}
\caption{Normalised sSFR distribution of  ``$\rm H\alpha$ excess'' galaxies at $3.9 \leq z \leq 4.9$. The main sequence and starburst galaxies are clearly separated at these redshifts. Here we denominate {\em `star formation valley'}  to the region empirically determined to be at $-8.05 \leq \rm log_{10}(sSFR) \leq -7.60$. This is analogous to the `green valley' between the main sequence and passive regime, which is observed at lower redshifts.} \label{fig_fracssfr}
}
\end{figure}

Fig.~\ref{fig_sfrvsstm} shows that the ``$\rm H\alpha$ excess'' galaxies form two distinct clouds on the SFR-$\rm M^\ast$ and sSFR-$\rm M^\ast$  planes. We identify these clouds as the so-called main-sequence of star formation \citep[e.g.,][]{noe07,rod10}  and a starburst cloud, which appears to be more prominent than what is observed at any lower redshift. The two cloud separation becomes completely evident when we plot the fraction of ``$\rm H\alpha$ excess'' galaxies versus sSFR (Fig.~\ref{fig_fracssfr}). From this Figure we can empirically derive that main-sequence galaxies are those with $\rm log_{10} sSFR \lsim-8.05$, while starburst galaxies have $\rm log_{10} sSFR \gsim-7.60$. 

In addition, there is an sSFR local minimum at $-8.05 \lsim \rm sSFR \lsim -7.60$ that here we denominate the {\em star formation valley} and likely contains galaxies in transition between the two star formation modes. This is analogous to the so-called `green valley' which in the literature refers to galaxies in transition from the main sequence to the passive regime \citep[e.g.,][]{bal11,zeh11,ren15}.

On the sSFR-$\rm M^\ast$ plot in Fig.~\ref{fig_sfrvsstm} we indicate upper limits to the median sSFR per stellar mass bin, obtained from {\em all} the galaxies at $3.9 \leq z \leq 4.9$. For the ``$\rm H\alpha$ excess''  galaxies we adopted the derived sSFR, while for the other galaxies we considered sSFR upper limits derived from the line complex EW (and corresponding $\rm H\alpha$ flux) upper limits. In that plot we clearly see that the upper limits to the median sSFR  decrease with increasing stellar mass in the stellar mass range $\rm 9.2 \leq log_{10} (M^\ast) \leq 10.8$. This trend is consistent with what is observed at lower $z$ \citep[e.g., ][]{kar11}.

We fitted the SFR-$\rm M^\ast$ relation in the two clouds formed by the ``$\rm H\alpha$ excess'' galaxies in  Fig.~\ref{fig_sfrvsstm} (left) with simple linear regressions and obtained: 
\vspace{-0.2cm}

\begin{equation}
\rm log_{10} (SFR)= (0.69 \pm 0.01) \times log_{10} (M^\ast)  - (5.44\pm0.13)
\label{eq-ms}
\end{equation}

\noindent for the main sequence and 
\vspace{-0.2cm}

\begin{equation}
\rm log_{10} (SFR)= (0.89 \pm 0.02) \times log_{10} (M^\ast)  - (6.18^{+0.16}_{-0.15})
\end{equation}

\noindent for the starburst cloud. In these relations, we have obtained the error bars on the slopes and intercepts through Markov Chain Monte Carlo (MCMC) fittings, assuming a crude 30\% average error on both the SFR and stellar mass. 

\begin{figure}[]
\center{
\includegraphics[width=1.1\linewidth, keepaspectratio]{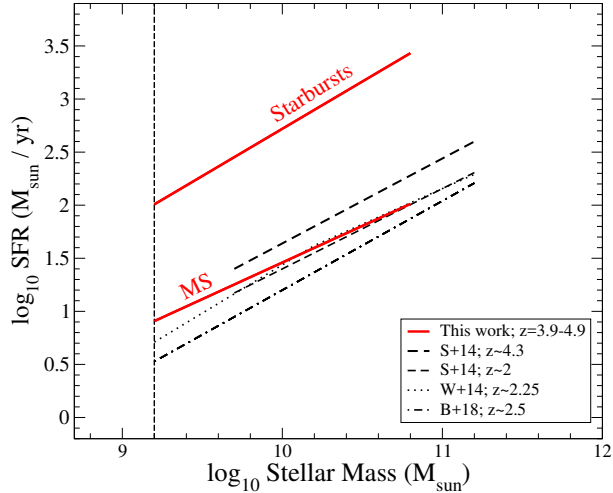}
\caption{Comparison of our ``$\rm H\alpha$ excess'' galaxy main sequence  with main-sequence determinations from the literature: S+14 \citep{spe14}; W+14 \citep{whi14} and B+18 \citep{bis18}, at different redshifts. Among these, only the \citet{bis18} main-sequence has been obtained after a starburst segregation as we have performed here. We also show our starburst cloud fitting for reference.} \label{fig_compms}
}
\end{figure}

The ``$\rm H\alpha$ excess'' galaxies identified as starbursts here constitute all the starbursts in our entire SMUVS sample at $3.9 \leq z \leq 4.9$, as virtually no SFR upper limit (corresponding to the non-``$\rm H\alpha$ excess'' galaxies) lies on the starburst cloud in Fig.~\ref{fig_sfrvsstm}. Instead, the main sequence defined by the ``$\rm H\alpha$ excess'' galaxies and fitted by Eq.~(\ref{eq-ms}) is not the complete main sequence at those redshifts. Therefore, we cannot use Eq.~(\ref{eq-ms}) along with analogous relations from the literature at lower redshifts to draw conclusions regarding the redshift evolution of the galaxy main sequence. Nonetheless, to put our results in context, in Fig.~\ref{fig_compms} we show our separate main sequence and starburst cloud fittings at $z=3.9-4.9$, along with the median main-sequence determination by \citet{spe14} at similar redshifts, which is given by $\rm log_{10} (SFR) \approx 0.80 \times log_{10}(M^\ast) - 6.36$ at  $z\sim4.3$. For reference, we also show some recent main-sequence determinations at $2<z<3$ \citep{spe14,whi14,bis18}. Among these literature results, only the \citet{bis18} curve has been obtained after performing a starburst/main-sequence separation as we do here. The others fit all star-forming galaxies onto a single sequence.

The analysis of the different curves in Fig.~\ref{fig_compms} tells us the following: firstly, the \citet{spe14} main-sequence curve at $z\sim 4.3$ is above our own, even when ours is an upper limit to the main-sequence which is biased for fitting only the ``$\rm H\alpha$ excess'' galaxies. This is simply because \citet{spe14} do not separate main sequence and starburst galaxies as we do, but rather fit a single relation for all star-forming galaxies. Secondly, our ``$\rm H\alpha$ excess'' galaxy main sequence broadly coincides with the main sequences by \citet{spe14} and \citet{whi14} at $z\sim2$.  We caution the reader against a wrong interpretation: this coincidence is just fortuitous, as our own main-sequence is an upper limit and the \citet{spe14} and \citet{whi14} main sequences include all star-forming galaxies, so any direct comparison could be misleading.
Our main-sequence determination can be more fairly compared with that of \citet{bis18} at $z\sim2.5$, which has been obtained after a starburst segregation. Taking into account that our own main sequence at $z=3.9-4.9$ is an upper limit to the true main sequence at these redshifts, we can conclude that there is probably little or no evolution of this sequence between $z\sim2.5$ and $z=3.9-4.9$.

\subsubsection{The origin of the sSFR bimodality}
\label{sec:origbim}

A priori it may seem surprising that the bimodality displayed by the $\rm H\alpha$ emitters on the SFR-$\rm M^\ast$ and sSFR-$\rm M^\ast$ planes is not observed in the rest EW versus stellar mass diagram (Fig.~\ref{fig_ewvsstm}, left). \citet{fum12} derived that the sSFR of an $\rm H\alpha$ emitter is proportional to the ratio between the line rest EW and the $R$-band mass-to-light ratio, i.e., $\rm sSFR \propto EW / (M^\ast/L_R)$, where $\rm L_R$ is the $R$-band luminosity. Therefore, one could reasonably expect that the distribution of $\rm M^\ast/L_R$ ratios is different for main sequence and starburst $\rm H\alpha$ emitters, and that this difference is to some extent responsible for the bimodal sSFR behaviour.

\begin{figure}[]
\center{
\includegraphics[width=1.1\linewidth, keepaspectratio]{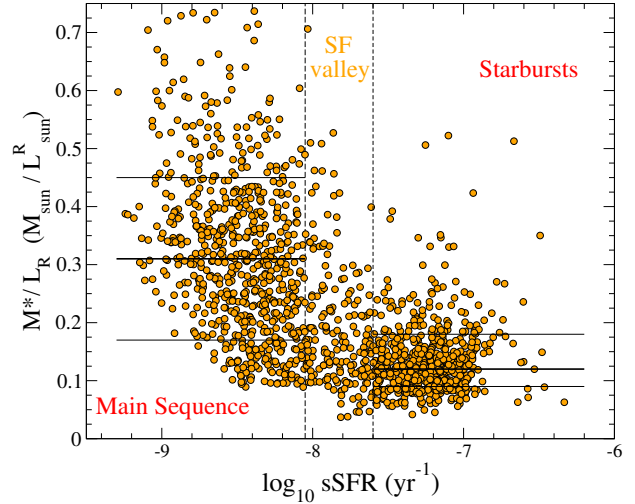}
\caption{$R$-band mass-to-light ratios versus sSFR for our ``$\rm H\alpha$ excess'' galaxies with stellar mass $9.2 \leq \rm log_{10} (M^\ast) \leq 10.8$. The vertical lines indicate our empirical separation between main sequence and starburst galaxies, as well as the star-formation (SF) valley, according to their sSFR (see Fig~\ref{fig_fracssfr}). The horizontal lines in each of the main-sequence/starburst regions indicate the 16th, 50th and 84th percentiles of the $\rm M^\ast/L_R$ distributions.} \label{fig_mlr}
}
\end{figure}

Figure \ref{fig_mlr} shows the $R$-band mass-to-light $\rm M^\ast/L_R$ ratio versus sSFR for our $\rm H\alpha$ emitters with stellar mass $9.2 \leq \rm log_{10} (M^\ast) \leq 10.8$. The $R$-band luminosities $\rm L_R$ considered here correspond to continuum luminosities, and have been obtained from the best-fit templates in LEPHARE's run excluding the $3.6 \, \rm \mu m$ photometry. Although there is no bimodality in the $\rm M^\ast/L_R$ ratios, a decreasing trend of these values with increasing sSFR is evident, with starburst galaxies having a tight distribution of small $\rm M^\ast/L_R$ values. For the starbursts, we find a median $\rm M^\ast/L_R=0.12$, with 16th-84th percentiles of 0.09-0.18. Instead, main-sequence $\rm H\alpha$ emitters typically have larger $\rm M^\ast/L_R$, following a wide distribution: the median is $\rm M^\ast/L_R=0.31$ and the 16th-84th percentiles are 0.17-0.45.

\begin{figure}[]
\center{
\includegraphics[width=1.1\linewidth, keepaspectratio]{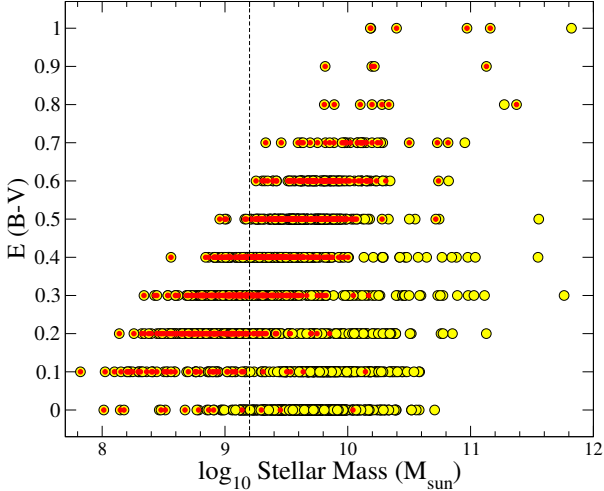}
\caption{Colour excess versus stellar mass for the ``$\rm H\alpha$ excess'' galaxies. Datapoints corresponding to our identified starbursts are highlighted with red dots.  The vertical line indicates the SMUVS 50\%  stellar-mass completeness at  $3.9 \leq z \leq 4.9$.} \label{fig_ebv}
}
\end{figure}

The small $\rm M^\ast/L_R$ ratios for starbursts are mainly the consequence of their systematic young ages, which typically are a few $\times 10^7 \, \rm yr$, according to their best SED fitting templates. However, starbursts are not the only galaxies in our sample which are characterised by such young ages: about 20\% of the ``non-$\rm H\alpha$ excess'' galaxies with stellar mass $9.2 \leq \rm log_{10} (M^\ast) \leq 10.8$ at $3.9 \leq z \leq 4.9$ have equally young best-fit ages in their SED fitting. Besides, starbursts are also characterised by having the largest extinction values among the ``$\rm H\alpha$ excess'' galaxies at fixed stellar mass, but there is no bimodality in the colour excess $\rm E(B-V)$ distribution, as can be seen in Fig.~\ref{fig_ebv}.  All these facts indicate that the {\em sSFR bimodality for the $H\alpha$ emitters does not arise as a consequence of a single galaxy property}, but is rather the consequence of the existence of a galaxy population (i.e., the starbursts) with a particular combination of properties: large $\rm H\alpha$ EW, young ages, and mostly high dust extinctions.

Although dust extinction and age can be degenerate in the SED fitting, this does not appear to be a significant problem for our starburst galaxies. If we analyze the probability density distribution as a function of age, marginalised over all other variables, we find that $\sim$90\% of the starbursts have 84th percentiles at ages $< 10^8 \,\rm yr$, i.e., they are truly galaxies dominated by young stellar populations, in contrast to the main-sequence $\rm H\alpha$ emitters, whose ages are mostly $> 10^8 \,\rm yr$.

\subsubsection{The importance of starburst galaxies at $z\sim4-5$}

The existence of different regions on the  SFR-$\rm M^\ast$ plane, corresponding to different modes of star formation,  has been analyzed in the literature at different redshifts, from the local Universe \citep{ren15} to $z\sim3$ \citep[e.g.,][]{san09,kaj10,wuy11,ilb15}. Here we show, for the first time, the existence of a prominent starburst sequence along with the main sequence on the SFR-$\rm M^\ast$ plane at $3.9 \leq z \leq 4.9$.

Starbursts constitute a small fraction (15\%) of {\em all} the SMUVS galaxies at $3.9 \leq z \leq 4.9$, but the percentage that we find here is significantly higher than the percentages found at $z\sim2-3$ \citep[e.g.,][]{rod11,sar12,sch15}. This is the case in spite of defining starbursts with an sSFR cut which is quite higher than the value $\rm log_{10} (sSFR) \gsim -8.1$ adopted by \citet{rod11} at $z\sim2$. At these lower redshifts, the percentage of galaxies with $\rm log_{10} (M^\ast) > 9$ and $\rm log_{10} (sSFR) > -7.60$ is negligible. 

Starbursts and main-sequence galaxies are similarly important in number among our  ``$\rm H\alpha$ excess'' galaxies with $\rm 9.2 \leq log_{10} (M^\ast) \leq 10.8$ at $3.9 \leq z \leq 4.9$. However, since starbursts make only $\sim 15\%$ of all the SMUVS galaxies in these stellar mass and redshift bins, the median sSFR upper limits shown in Fig.~\ref{fig_sfrvsstm} (right) lie all on the star-formation main sequence. 
 
\begin{figure}[]
\center{
\includegraphics[width=1.1\linewidth, keepaspectratio]{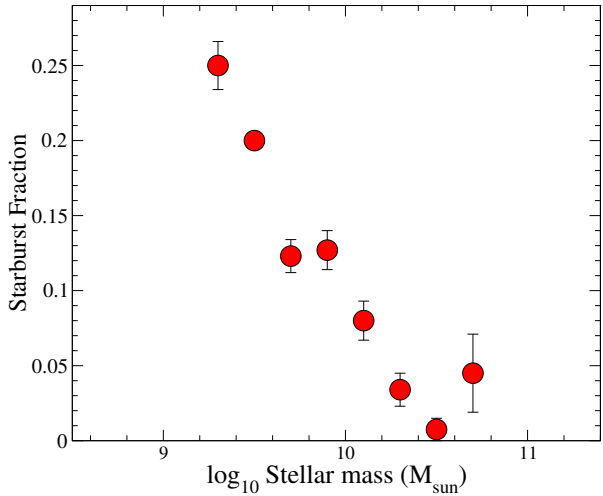}
\caption{Fraction of starbursts (defined as galaxies with $\rm log_{10}(sSFR)>-7.60$) in different stellar mass bins among all the SMUVS galaxies at $3.9 \leq z \leq 4.9$.} \label{fig_fracstb}
}
\end{figure}

Figure~\ref{fig_fracstb} shows how the fraction of starburst galaxies at $3.9 \leq z \leq 4.9$ varies with stellar mass. These fractions range from $\sim 0.25$ at $\rm log_{10} (M^\ast) \sim 9.3$ to $<0.05$ at $\rm log_{10} (M^\ast) > 10.2$, showing that the starburst phenomenon is much more common among intermediate-mass than massive galaxies at $z\sim4-5$, consistently to what is predicted by theoretical galaxy models \citep[e.g.,][]{cow17} and observed at lower redshifts \citep[e.g.,][]{bis18}.

It is interesting to do a back-of-the-envelope calculation of how much stellar mass these starbursts could accumulate at these redshifts. A $\rm log_{10}(sSFR)=-7.2$ value implies a stellar-mass doubling time of $\rm 1/sSFR \approx 1.6 \times 10^7 \, yr = 0.016  \, Gyr$. The elapsed time between $z=4.9$ and $z=3.9$ is 0.379~Gyr. Besides, from Fig.~\ref{fig_fracstb} we see that the fraction of galaxies with $\rm log_{10} (M^\ast) \sim 9.5$ that are starbursts is 0.20.  So if we assume that all galaxies with similar stellar mass would spend a similar amount of time in the starburst phase, then the  $\rm log_{10} (M^\ast) \sim 9.5$ galaxies would spend a fifth of the above-mentioned elapsed time as starbursts, i.e., $\rm \sim 0.076 \, Gyr$, which is almost five times the stellar-mass doubling time.  Therefore, assuming a recycled fraction of 50\%, we infer that these galaxies could grow their stellar mass by a factor of $\sim 2.5$ at those redshifts. This means that a typical  $\rm M^\ast \sim 3 \times 10^9 \, M_\odot$ galaxy at $3.9 \leq z \leq 4.9$ would become a galaxy with $\rm M^\ast \sim 8 \times 10^{9} \, M_\odot$ at $z<3.9$. This crude estimate shows that the large sSFR values derived here are consistent with a rapid, but plausible, growth of intermediate-mass galaxies happening $\rm \sim 1.5 \, Gyr$ after the Big Bang.

\begin{figure*}[]
\center{
\includegraphics[width=0.7\linewidth, keepaspectratio]{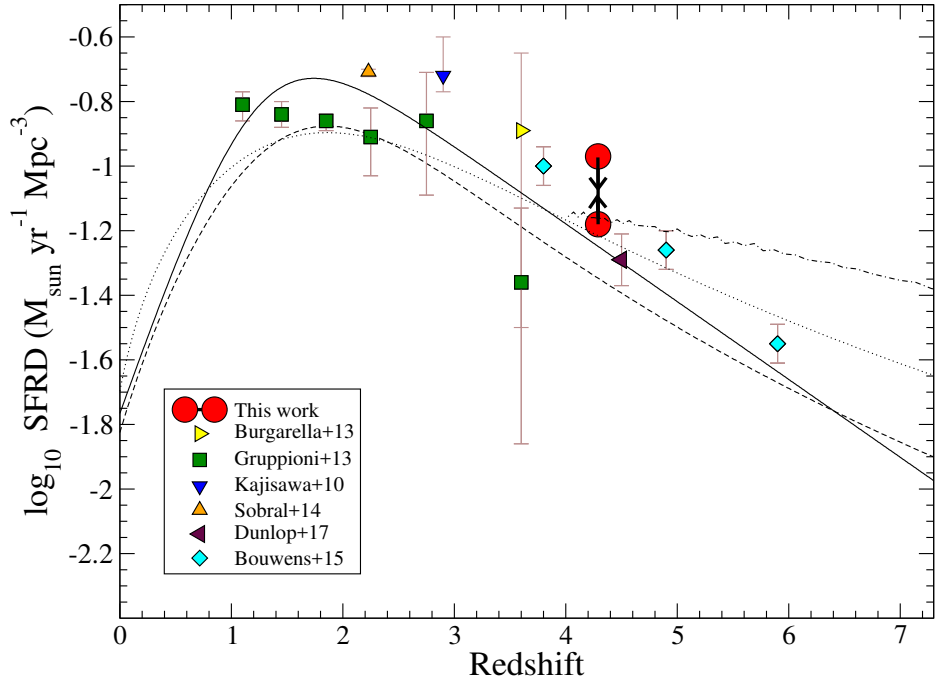}
\caption{Cosmic star formation rate density versus redshift. The large red circles at $z=4.29$ indicate our own lower and upper limit determinations. To derive the SFRD lower limit, we considered only the SFR of  ``$\rm H\alpha$ excess''  galaxies with  $\rm log_{10}(M^\ast) \leq 10.8$. To derive the upper limit,  we took into account the SFR of all SMUVS galaxies with $9.2 \leq \rm log_{10}(M^\ast) \leq 10.8$  at $3.9 \leq z \leq 4.9$ (exact values for the ``$\rm H\alpha$ excess''  galaxies and upper limits for all the others), and applied a correction factor for lower stellar masses inferred from the GSMF extrapolation at those redshifts (see text). In both cases, only galaxies with $\rm log_{10}(M^\ast) \leq 10.8$  have been taken into account to minimize any possible AGN contamination. Other symbols show recent SFRD determinations from the literature, based on different SFR tracers. The different curves correspond to either data-compilation best fits or theoretical predictions from the literature. {\em Solid line:}  \citet{beh13}; {\em dashed-dotted line:} \citet{day14}; {\em dashed line:}  \citet{mad14}; {\em dotted line:} \citet{kha15}.  All SFRD values in this figure correspond to a \citet{sal55} IMF over stellar masses $\rm (0.1-100) \, \rm M_\odot$. This is the only figure in this paper in which this IMF is adopted, only for the purpose of facilitating the comparison with the literature. \label{fig_sfrdvsz}
}}
\end{figure*}

\subsection{The inferred cosmic star formation rate density at $\lowercase{z}=3.9-4.9$\label{subsec:sfrd}}

We can use our derived SFR to obtain an estimate of the cosmic SFR density (SFRD) at $3.9 \leq z \leq 4.9$. Considering only the ``$\rm H\alpha$ excess''  galaxies gives us a lower limit to the SFRD. Instead, if we added the SFR contributions of all the other SMUVS galaxies at these redshifts, for which we only have SFR upper limits, we could estimate an upper limit to the SFRD. However, we recall that SMUVS significantly loses completeness at stellar masses $\rm log_{10}(M^\ast)<9.2$, so even if we apply completeness corrections, we can only extrapolate the missing galaxy SFRs based on the galaxies that we detect. Therefore, estimating a more secure upper limit requires accounting somehow for the galaxies that we do not see. This can be done, at least in a crude manner, taking into account the faint-end slope of the galaxy stellar mass function (GSMF) at those redshifts and assuming that the SFR versus stellar mass trends shown in Fig.~\ref{fig_sfrvsstm} can be extrapolated to lower stellar masses.

Figure~\ref{fig_sfrdvsz} shows the redshift evolution of the SFRD in the so-called Lilly-Madau diagram \citep{lil96,mad96}. In this plot we have included our derived lower and upper limits to the SFRD at a median redshift $\langle z \rangle=4.29$, as well as a compilation of values recently reported in the literature, which are based on different galaxy surveys and individual galaxy SFRs from a variety of SFR tracers: UV fluxes \citep{bou15}; spectral line emission from narrow-band surveys \citep{sob14}; far-IR fluxes \citep{gru13}; and a combination of UV and IR fluxes \citep{kaj10,bur13,dun17}. Multiple other previous works have presented SFRD estimates at high $z$, especially based on UV fluxes. However, UV-based SFRs are much more affected by dust-extinction corrections than those based on any other tracer, making the resulting SFRD values particularly uncertain at least to $z\sim6$ \citep[see e.g.,][]{cas14}.

In the derivation of our own SFRD, we have explicitly excluded all galaxies with $\rm log_{10}(M^\ast)>10.8$ to minimize any plausible AGN contamination (see discussion in Section~\ref{subsec:sfrstm}). By considering only the SFR of ``$\rm H\alpha$ excess'' galaxies we obtained a robust SFRD lower limit of $0.066 \, \rm  M_\odot \, yr^{-1} \, Mpc^{-3}$. To obtain an SFRD upper limit, we estimated separately the contributions of {\em all} galaxies with  $\rm 9.2 \leq log_{10} M^\ast \leq 10.8$ and $\rm 8 \leq log_{10} M^\ast < 9.2$ and added them up, as follows.  For the $\rm 9.2 \leq log_{10} M^\ast \leq 10.8$ objects we directly adopted our derived SFR  (``$\rm H\alpha$ excess'' galaxies) or SFR upper limits (for the other galaxies). To account for the contribution of lower mass galaxies we considered that the extrapolation of the GSMF at $z=4-5$ \citep{cap15,gra15} indicates that we should expect $\sim 5$ times more galaxies with $\rm 8 \leq log_{10} M^\ast < 9.2$ than with $\rm log_{10} M^\ast \geq 9.2$.  And according to Fig.~\ref{fig_sfrvsstm}, the $\rm 8 \leq log_{10} M^\ast < 9.2$ objects should have on average SFR that are about a dex lower than the more massive galaxies. Thus, the lower mass galaxies should roughly add $\sim 50\%$  to the SFRD value calculated based only on the $\rm 9.2 \leq log_{10} M^\ast \leq 10.8$ sources.  With all these considerations, we obtain an SFRD upper limit of $0.106 \, \rm  M_\odot \, yr^{-1} \, Mpc^{-3}$. All the SFRD quoted here and in  Fig.~\ref{fig_sfrdvsz} correspond to a \citet{sal55} IMF over stellar masses $\rm (0.1-100) \, \rm M_\odot$ to facilitate the comparison with the literature, in which this IMF is mostly used. Our resulting range of SFRD at $\langle z\rangle=4.29$ is in good agreement with the other recent observational determinations in the literature.

The different curves in Fig.~\ref{fig_sfrdvsz} show best fits to compilations of observational data \citep{beh13,mad14} and theoretical predictions \citep{day14,kha15}.  We see that the majority of the most recent SFRD estimates at high $z$, including our own,  are at least 0.10-0.15~dex above the best fits provided by \citet{mad14} and \citet{beh13}. This is probably because most of the observational works considered for these fits calculated SFRs based on UV fluxes, which are very sensitive to dust extinction corrections, as mentioned above. A revision of these corrections \citep{bou15} or, even better, a consideration of UV+IR data \citep{kaj10,bur13} or SFRs based on line emission like those we obtain here should provide better constraints to the cosmic SFRDs at high $z$ at least to $z\sim6$. Dust extinction corrections are expected to be less critical at earlier cosmic times, but this will have to be confirmed in the future, with systematic IR galaxy surveys that can trace Balmer line emission in the early Universe, which will happen with the {\em James Webb Space Telescope (JWST)}.  The theoretical predictions appear to be in good agreement with our SFRD lower limit at $\langle z \rangle=4.29$, as well as some of the other observational data points.

The contribution of starburst galaxies to the total SFRD budget up to $z\sim3$ has been discussed in the literature \citep[e.g.,][]{rod11,sar12,sch15}. The derived contributions vary according to the selection effects and the starburst definition adopted by different groups, but in all cases it has been found that starbursts make $\lsim 15\%$ of the total SFRD.  As we discussed before, all these studies have been based on the analysis of $\rm M^\ast \gsim 10^{10} \, \rm M_\odot$ galaxies. Here, we obtain that the ``$\rm H\alpha$ excess'' galaxies defined as starbursts by $\rm log_{10}(sSFR) \gsim-7.60$ account for $>50\%$ of our upper limit to the SFRD (and $84\%$ of our lower limit based only on the ``$\rm H\alpha$ excess'' galaxies) at $3.9 \leq z \leq 4.9$. These percentages are substantially higher than all previous determinations. The key reason for this higher percentage is that here we are including galaxies down to lower stellar masses, which are much more numerous than massive galaxies and have a higher fraction of starbursts among them.

\begin{figure*}[]
\center{
\includegraphics[width=0.7\linewidth, keepaspectratio]{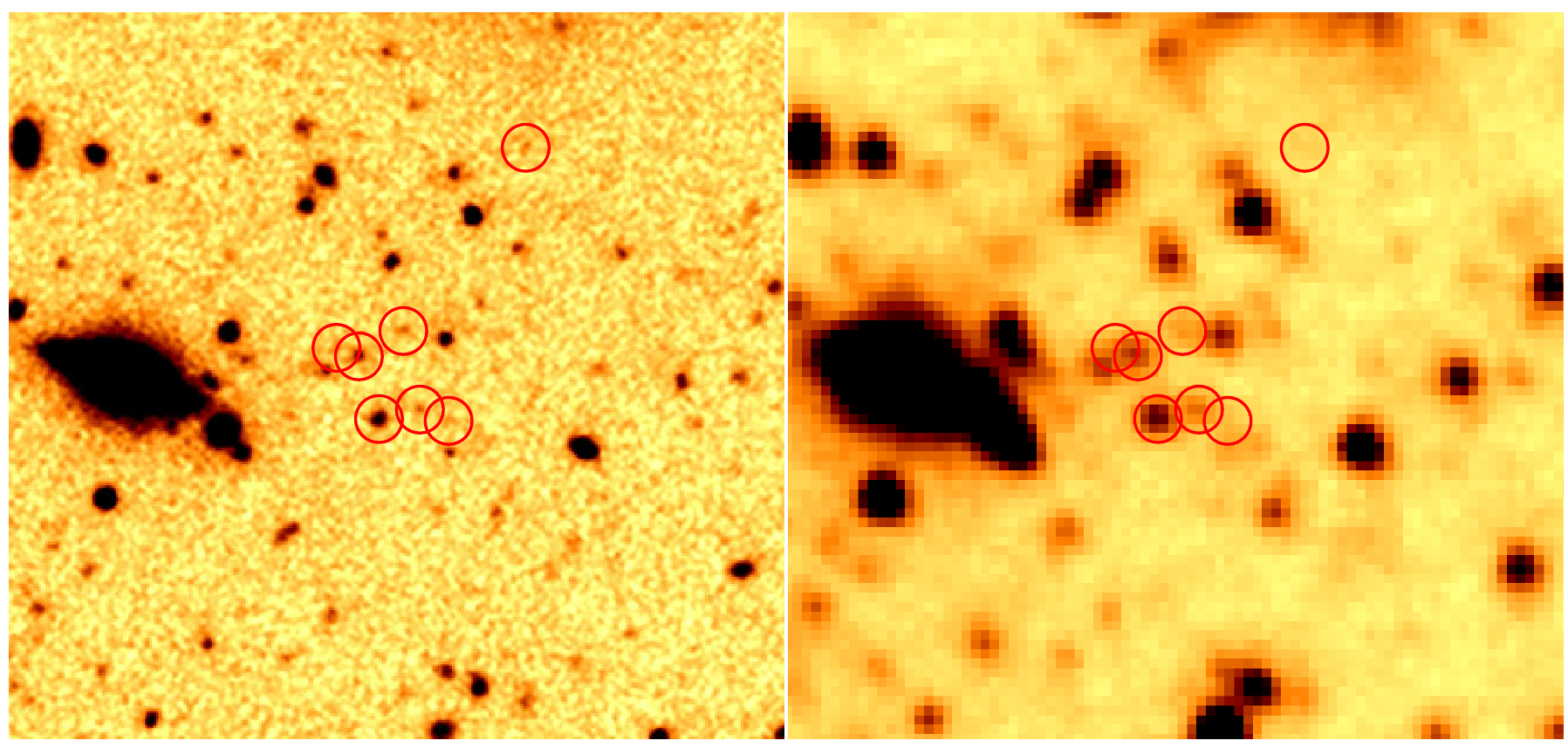}
\caption{A $>50 \sigma$ overdensity of SMUVS galaxies at $z\approx 4.2$. {\em Left:} UltraVISTA $HK_s$ mosaic; {\em right:} SMUVS $3.6 \, \rm \mu m$ mosaic. In addition to the six galaxies making the unusually high overdensity, a seventh galaxy with similar redshift is present only $\sim0.3$~arcmin apart. The properties of individual galaxies are listed in Table~\ref{tab_protocl}. Each image corresponds to an area of $\sim 0.8 \times 0.8 \, \rm arcmin^2$. \label{fig_protocl}
}}
\end{figure*}

\begin{deluxetable*}{rcccrrrc}[b!]
\tablecaption{Properties of individual sources in the $z\approx 4.2$ galaxy overdensity. \label{tab_protocl}}
\tablecolumns{8}
\tablenum{1}
\tablewidth{0pt}
\tablehead{
\colhead{ID} &
\colhead{RA (J2000)} &
\colhead{DEC (J2000)} &
\colhead{$z_{\rm phot}$} &
\colhead{$\rm log_{10}M^\ast \, (M_\odot)$} &
\colhead{SFR~$\rm (M_\odot / yr)$} & 
\colhead{ID (L+16)}  &
\colhead{$z_{\rm phot}$ (L+16)}
}
\startdata
SM2\_103948 & 10:00:37.43 & +02:25:40.25 &  4.83$\pm0.10$  &  9.71$^{+0.08}_{-0.08}$ & 36.9$\pm$7.2 & -- & -- \\
SM2\_103968 & 10:00:37.55 & +02:25:40.97 &  4.57$\pm0.10$  & 10.22$^{+0.08}_{-0.50}$ &  8.6$\pm$8.5  & --  & -- \\
SM2\_104007 & 10:00:37.73 & +02:25:40.37 &  4.07$\pm0.10$  &  9.80$^{+0.08}_{-0.09}$ & 53.3$\pm$13.9 & 743559 & 4.05\\
SM2\_104067 & 10:00:37.91 & +02:25:44.92 &  4.22$\pm0.10$  &  9.06$^{+0.08}_{-0.08}$ &  $<9.8$  & -- & -- \\
SM2\_104073 & 10:00:37.81 & +02:25:44.37 &  3.93$\pm0.10$  &  9.02$^{+0.08}_{-0.08}$ & 83.5$\pm$9.1 & 744162 & 4.10\\
SM2\_104100 & 10:00:37.62 & +02:25:46.00 &  3.97$\pm0.10$  &  8.77$^{+0.08}_{-0.09}$ & 28.2$\pm$3.2 & 744344 & 4.13 \\
\hline
SM2\_104510 & 10:00:37.10  &+02:25:57.71 &  4.26$^{+0.15}_{-3.81}$  &  7.79$^{+0.11}_{-0.10}$ & $<0.5$ & 746450 & 0.36 \\
\enddata
\tablecomments{Source coordinates correspond to the $HK_s$ mosaic. The SFR values are based on the inferred $\rm H\alpha$ rest EWs and fluxes derived from the $3.6 \, \rm  \mu m$ photometry. The last two columns contain ID and photometric redshifts from the COSMOS 2015 catalogue by Laigle et al.~(2016; L+16). The last row corresponds to a galaxy outside the main overdensity region, but only $\sim 0.3 \, \rm arcmin$ apart from it.}
\end{deluxetable*}

\subsection{A $>50 \sigma$ overdensity of $3.9 \leq z \leq 4.9$ galaxies \label{subsec:protoclust}}

Within our analysis of SMUVS galaxies at  $3.9 \leq z \leq 4.9$, we find that six of these sources lie in a very small region of only $\sim 0.20 \times 0.20$~arcmin$^2$, with a median redshift $z\approx 4.2$. We have a total of 5925 SMUVS galaxies at those redshifts over a total area of 0.66~deg$^2$, which makes for an average surface density of $\sim$2.5~arcmin$^{-2}$. Therefore, the six galaxies concentrated in the tiny  $\sim 0.20 \times 0.20$~arcmin$^2$ area constitute a remarkable $>50 \sigma$ overdensity of sources at $3.9 \leq z \leq 4.9$. 

Figure~\ref{fig_protocl} shows the UltraVISTA $HKs$ stack and SMUVS $3.6 \, \rm \mu m$ images of this galaxy overdensity. The properties of the six individual galaxies, as well as those of another galaxy at a similar (albeit less secure) redshift, which lies only $\sim$0.3~arcmin apart, are given in Table~\ref{tab_protocl}.

In the main overdensity region, three out of six galaxies have redshifts $3.93 \leq z \leq 4.07$. Within the error bars, this suggests that these galaxies are likely part of the same bound structure. These three galaxies are also present in the COSMOS2015 catalogue by \citet{lai16}, whose photometry and redshifts have been independently determined based on a previous release of the UltraVISTA data (DR2) and shallower {\em Spitzer} data (see Table~\ref{tab_protocl}). Their photometric redshifts are consistent with ours within the error bars. These independent results support our claim that a $z\sim 4$ overdensity is present in that region of the sky. The other three galaxies that we find in the overdesity region are at higher redshifts $4.22 \leq z \leq 4.83$, so spectroscopic data are needed to determine whether they belong to the same structure or not.

Among the six galaxies in the overdensity,  five are classified as ``$\rm H\alpha$ excess'' sources, although for one of them this classification is marginal. The other four have SFRs corresponding to several tens $\rm M_\odot / yr$. Note, however, that only one of them (ID SM2\_104073) is a starburst galaxy, while the others are main sequence galaxies. All these galaxies have intermediate stellar masses, with only one of them having $\rm log_{10}M^\ast>10$.

\section{Discussion and Conclusions} \label{sec:concl}

In this paper we have studied a sample of 5925 galaxies at $3.9 \leq z \leq 4.9$ from the SMUVS survey over 0.66~deg$^2$ of the COSMOS field. We have analyzed the presence of flux excess in the IRAC $3.6 \, \rm \mu m$ band to identify ``$\rm H\alpha$ excess''  galaxies at these redshifts.  From the inferred $\rm H\alpha$ EW we have derived SFR and sSFR for these galaxies and obtain upper limits for all the other galaxies at these redshifts. 

The incidence of  ``$\rm H\alpha$ excess''  galaxies in our $3.9 \leq z \leq 4.9$  sample decreases from stellar masses $\rm log_{10} (M^\ast) \sim 9.2$ through $\rm log_{10} (M^\ast) \sim 10.8$, indicating that intense star formation is relatively more important among intermediate-mass than massive galaxies at these redshifts.  At higher stellar masses, surprisingly,  we see a reversing trend which seems difficult to explain by simply invoking enhanced star formation for the most massive systems. As a large fraction of the most massive galaxies with $3.6 \, \rm \mu m$ flux excess at $3.9 \leq z \leq 4.9$ are $24 \, \rm \mu m$-detected, we suggest that AGN activity could at least be partly responsible for this flux excess. This statement is supported by a spectroscopically confirmed AGN among the most massive ``$\rm H\alpha$ excess'' galaxies. 

The most important result in this paper is that we found that the ``$\rm H\alpha$ excess''  galaxies form two distinct sequences on the SFR-$\rm M^\ast$ plane, which we recognise as the so-called star-formation main sequence and a starburst sequence. Although the exact location of galaxies on these planes depends on the SED fitting assumptions, we have shown that the sSFR bimodality is independent of the assumed metallicities and dust reddening law, under reasonable assumptions.  Starburst galaxies are characterised by a unique combination of large $\rm H\alpha$ EW, young ages and mostly high dust extinctions, which result in very high sSFR.

This suggests the existence of two distinct star formation modes,  one corresponding to secular baryonic accretion within dark matter haloes (giving rise to the main sequence), and another one much more effective for rapid galaxy growth, possibly linked to galaxy mergers or external perturbations that produce the starburst phase. A fundamental intrinsic assumption in all these results is that the same conversion from gas into stars given by Kennicutt's law (equation~\ref{eq-kenn}) is valid for all star-forming galaxies at high $z$. Demonstrating this general validity is extremely difficult, but recent studies of the link between gas content and star formation in high-$z$ galaxies suggest that a universal conversion may indeed hold \citep[e.g., ][]{sco16}.

The star-formation main sequence at different redshifts has been analyzed from multiple observational datasets \citep[e.g.,][]{bri04,noe07,rod10,elb11,spe14,whi14,tas15}. Theoretical studies have attempted to explain the main sequence as a mere consequence of gas accretion within dark matter haloes \citep[e.g.,][]{som08,lu14,cou15}, but have difficulty in reproducing the evolution of its normalization with cosmic time \citep{dut10,gen14,fur15,spa15} -- see \citet{mit14} for a thorough discussion of this problem. Other groups have investigated the role of galaxy-black hole coevolution in shaping the star formation main sequence \citep[e.g.,][]{man16,kav17}.  In any case,  a complete theoretical explanation of this relation between stellar mass and instantaneous star formation rate is still missing.

In contrast to the widely discussed main sequence,  most observational studies have failed to recognise the presence of a separate starburst cloud on the SFR-$\rm M^\ast$ plane (its existence has been suggested by \citet{cas16}, but these authors referred to it as a ``second main sequence''). The failure to recognise the starburst cloud is mainly due to the relatively small galaxy samples analyzed in most of the literature, with starburst galaxies typically identified and studied only among massive galaxies  \citep{rod11,sar12}. Here we show, for the first time, that starburst galaxies make a clearly distinct sequence on the SFR-$\rm M^\ast$ plane and constitute a significant percentage ($\sim 15\%$) of {\em all} galaxies with $\rm 9.2 \leq log_{10} M^\ast \leq 10.8$  at  $3.9 \leq z \leq 4.9$.  Only a slightly smaller percentage is obtained at $z\sim 2-3$ when a broad dynamic range in stellar mass is considered, albeit with significantly lower sSFR \citep{bis18}.  We find that the fraction of starburst galaxies has a strong dependence on stellar mass, varying between $\sim 0.25$ at $\rm log_{10} (M^\ast) \sim 9.3$ to $<0.05$ at $\rm log_{10} (M^\ast) > 10.2$. This is also expected from galaxy formation models \citep[e.g.,][]{vog14}.

The starburst galaxies in our sample are characterised by $\rm log_{10}(sSFR)>-7.60$. At redshifts $z \lsim 4$, these very high sSFR values are very unusual and have only been found to be common among low stellar-mass $\rm M^\ast \lsim 10^8 \, \rm M_\odot$ young galaxies -- see, e.g., Fig.~11 in \citet{kar17}, and also \citet{amo17,van17}. Here we show that these high sSFRs are relatively common among star forming, intermediate-mass galaxies at $3.9 \leq z \leq 4.9$.

Another key result of this paper is that starbursts make for at least $50\%$ of the total SFRD budget at $z\sim 4-5$. This percentage is substantially higher than any previous determination found in the literature. This is because previous starburst studies only analyzed galaxies with $\rm M^\ast \gsim 10^{10} \, \rm M_\odot$ (up to $z\sim3$). The inclusion of lower mass galaxies is essential to recognise the importance of the starburst phase and how much it contributes to the total cosmic SFRD, particularly at high redshifts.

We have also discovered an unusually high-significance galaxy overdensity at these high redshifts. A total of six galaxies in our high-$z$ sample reside in a tiny area of $0.20 \times 0.20$~arcmin$^2$, five of which correspond to ``$\rm H\alpha$ excess'' sources.  The spectroscopic confirmation of such an overdensity would unveil one of the most concentrated active sites of star formation known at $z\sim 4-5$.

All the results presented in this paper have exploited the unique combination of area and depth provided by the SMUVS survey at mid-IR wavelengths. This survey has provided us with an unprecedented level of statistics and dynamic range which are fundamental to reveal unknown aspects of galaxy evolution in the young Universe.

\acknowledgments

Based in part on observations carried out with the {\em Spitzer Space Telescope}, which is operated by the Jet Propulsion Laboratory, California Institute of Technology under a contract with NASA. Also based on data products from observations conducted with ESO Telescopes at the Paranal Observatory under ESO program ID 179.A-2005 and on data products produced by TERAPIX and the Cambridge Astronomy Survey Unit on behalf of the UltraVISTA consortium. Also based on observations carried out by NASA/ESA {\em Hubble Space Telescope}, obtained and archived at the Space Telescope Science Institute; and the Subaru Telescope, which is operated by the National Astronomical Observatory of Japan.  This research has made use of the NASA/IPAC Infrared Science Archive, which is operated by the Jet Propulsion Laboratory, California Institute of Technology, under contract with NASA.

We thank an anonymous referee for a constructive report. KIC, SD and WC  acknowledge funding from the European Research Council through the award of the Consolidator Grant ID 681627-BUILDUP.

%% To help institutions obtain information on the effectiveness of their 
%% telescopes the AAS Journals has created a group of keywords for telescope 
%% facilities.
%
%% Following the acknowledgments section, use the following syntax and the
%% \facility{} or \facilities{} macros to list the keywords of facilities used 
%% in the research for the paper.  Each keyword is check against the master 
%% list during copy editing.  Individual instruments can be provided in 
%% parentheses, after the keyword, but they are not verified.

\vspace{5mm}
\facilities{Spitzer, VISTA, Subaru}

%% Similar to \facility{}, there is the optional \software command to allow 
%% authors a place to specify which programs were used during the creation of 
%% the manusscript. Authors should list each code and include either a
%% citation or url to the code inside ()s when available.

\software{SExtractor, IRAF, LePhare}

%% Appendix material should be preceded with a single \appendix command.
%% There should be a \section command for each appendix. Mark appendix
%% subsections with the same markup you use in the main body of the paper.

%% Each Appendix (indicated with \section) will be lettered A, B, C, etc.
%% The equation counter will reset when it encounters the \appendix
%% command and will number appendix equations (A1), (A2), etc. The
%% Figure and Table counter will not reset.

\appendix

\section{Impact of considering two metallicities for the $\rm H\alpha$ excess galaxies}

\begin{figure*}[ht!]
\center{
\includegraphics[width=1\linewidth, keepaspectratio]{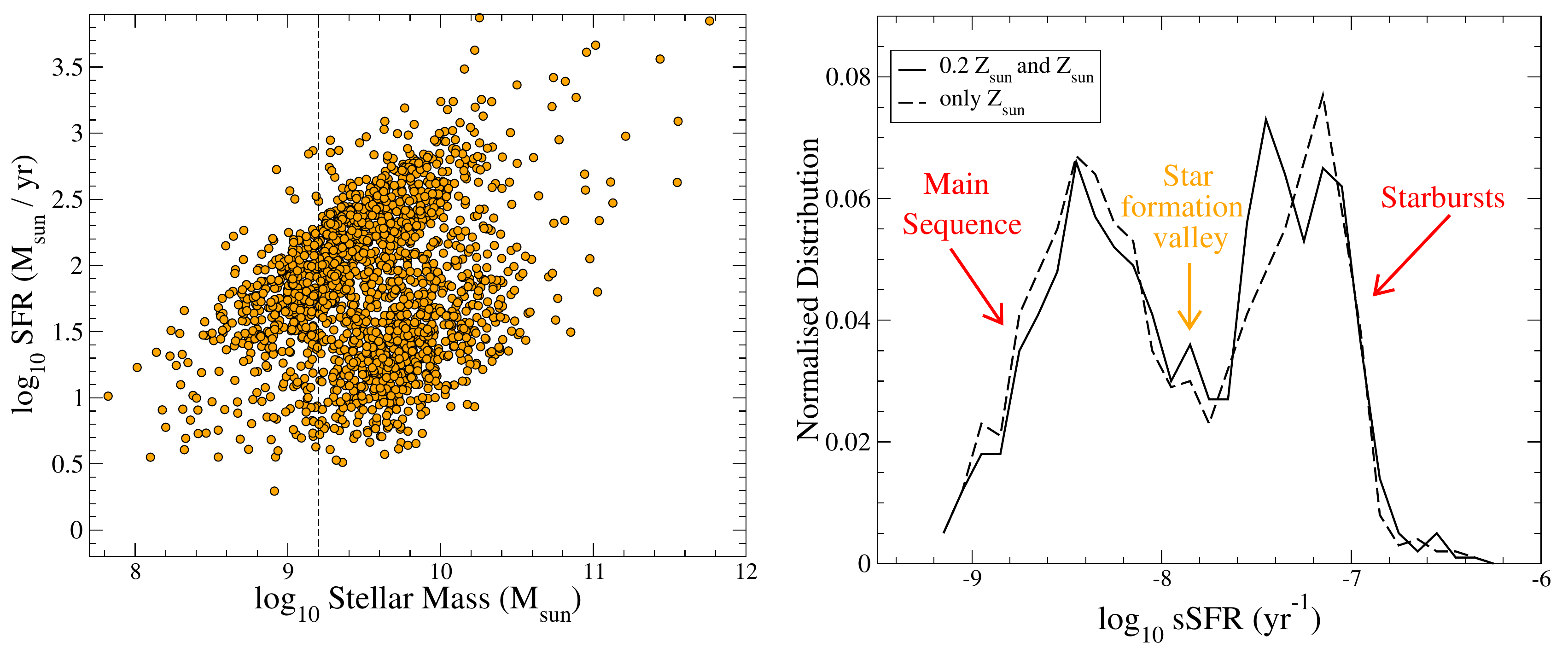}
\caption{{\em Left:} SFR based on the derived $\rm H \alpha$ luminosities for ``$\rm H\alpha$ excess'' galaxies  versus stellar mass at $3.9 \leq z \leq 4.9$, obtained after allowing for two possible metallicities ($\rm Z_\odot$ and $\rm 0.2 \, Z_\odot$) in the SED fitting.  {\em Right:} resulting sSFR distribution for the ``$\rm H\alpha$ excess'' galaxies with $9.2 \leq \rm log_{10} (M^\ast) \leq 10.8$.} \label{fig_twomet}
}
\end{figure*}

In this paper we have modelled the SEDs of our $3.9 \leq z \leq 4.9$ galaxies assuming a solar metallicity. Other authors have preferred to adopt a much lower metallicity ($\rm 0.2 \, Z_\odot$) to study ``$\rm H\alpha$ excess'' galaxies at similar redshifts \citep[e.g.,][]{smi16}. However, in a stellar-mass selected galaxy sample like SMUVS, the adoption of sub-sollar metallicities for all galaxies is likely not correct \citep[e.g.][]{sco16}. The most likely scenario is that our galaxies span a range of metallicities between a fraction of the solar and the entire solar values.

To test the impact of this effect on our results, we have re-run the SED fitting of our  ``$\rm H\alpha$ excess''  galaxies at $3.9 \leq z \leq 4.9$  allowing LePhare to use both templates with $\rm 0.2 \, Z_\odot$ and $\rm Z_\odot$ metallicities, covering the same grid for all the other parameter values (i.e., star formation histories, ages, extinctions).  By doing this, we found that: i) 96\% of all our original ``$\rm H\alpha$ excess''  galaxies are confirmed to be in the redshift range  $3.9 \leq z \leq 4.9$, and thus confirmed as $\rm H\alpha$ emitting sources; ii) about 34\% of the original ``$\rm H\alpha$ excess''  galaxies have a best-fit SED with $\rm 0.2 \, Z_\odot$, while $\sim 62\%$ have a best-fit SED with solar metallicity.

Taking into account these results, we re-computed the $\rm H\alpha$ rest EW and fluxes for the ``$\rm H\alpha$ excess''  galaxies that have a best-fit SED with $\rm 0.2 \, Z_\odot$ in the new LePhare run. At this low metallicity, the net $\rm H\alpha$ contribution to the total $3.6 \, \rm \mu m$ flux excess is given by \citep{and03}: $f(\rm H\alpha)$$= 0.81 f(\rm H\alpha + [NII]+[SII])$ and $f(\rm H\alpha)$$= 0.92 f(\rm H\alpha + [NII])$, which are valid at $3.9 \leq z \leq 4.8$ and  $4.8 < z \leq 4.9$, respectively. Finally, we derived the corresponding clean $\rm H\alpha$ luminosities and SFR in the same way as explained in Section~\S\ref{subsec:sfrstm}.

Figure~\ref{fig_twomet} (left) shows the resulting SFR-$\rm M^\ast$ plane for ``$\rm H\alpha$ excess''  galaxies when considering two metallicities for the SED fitting (we only plotted here the 96\% of the original ``$\rm H\alpha$ excess'' sources that stayed in the $3.9 \leq z \leq 4.9$  redshift range in the new SED-fitting run). Qualitatively, this plot looks similar to the analogous plot in Fig.~\ref{fig_sfrvsstm}, with two distinct galaxy star-formation main sequence and starburst cloud clearly visible. Fig.~\ref{fig_twomet} (right) shows the corresponding sSFR distribution compared to the original distribution derived adopting only solar metallicity  (see Fig.~\ref{fig_fracssfr}). This plot confirms the existence of two sSFR regimes, separated by a star formation valley corresponding to an underdensity of sources with $-8.05 \lsim \rm sSFR \lsim -7.60$. Therefore, we can conclude that all the main results in this paper are robust against our SED metallicity assumptions.

\section{Impact of considering a reddening law dependent on the UV slope $\beta$}

Another parameter that could potentially influence our results is the choice of dust reddening law in the SED fitting and recovery of intrinsic $\rm H\alpha$  luminosities. In this work we have adopted the \citet{cal00} reddening law for the SED fitting and applied the derived colour excess from the best-fit SED of each ``$\rm H\alpha$ excess''  galaxy to determine its intrinsic $\rm H\alpha$  flux. Here we investigate the impact of a different assumption, namely, that the internal extinction of each galaxy is directly related to its UV spectral slope $\beta$, which is defined as the slope in  $f_\lambda(\lambda) \propto \lambda^\beta$ for rest-frame $1500 < \lambda < 2500 \, \rm \AA$.

\begin{figure*}[ht!]
\center{
\includegraphics[width=1\linewidth, keepaspectratio]{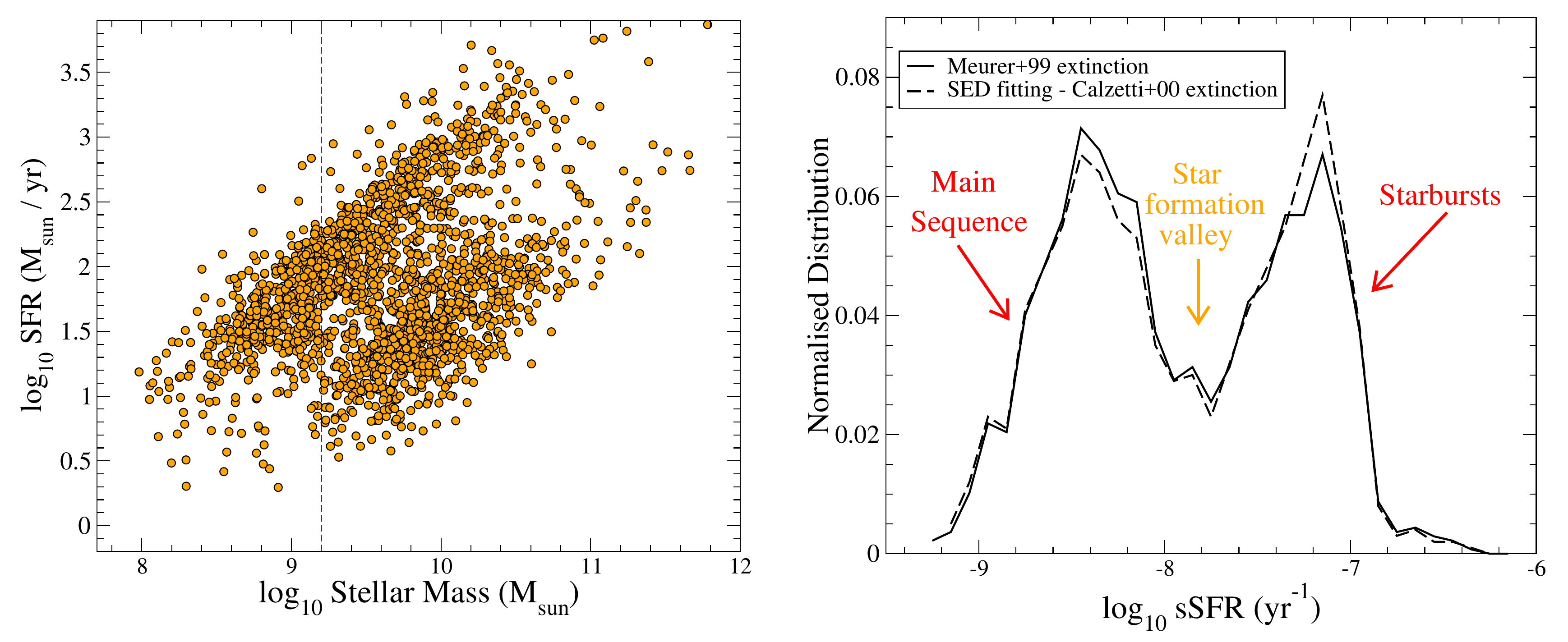}
\caption{{\em Left:} SFR based on the derived $\rm H \alpha$ luminosities for ``$\rm H\alpha$ excess'' galaxies  versus stellar mass at $3.9 \leq z \leq 4.9$, obtained after considering a dust extinction correction dependent on the rest UV slope of each galaxy, following the \citet{meu99} relation.  {\em Right:} resulting sSFR distribution for the ``$\rm H\alpha$ excess'' galaxies with $9.2 \leq \rm log_{10} (M^\ast) \leq 10.8$.} \label{fig_meurext}
}
\end{figure*}

We computed the UV slope $\beta$ of each of our galaxies by fitting this power-law functional form to the photometric fluxes tracing rest $1500 < \lambda < 2500 \, \rm \AA$, which correspond to about ten photometric bands in the COSMOS field \citep[as has been done by e.g.,][]{fud17}. Then we assumed that the internal dust extinction was given by the \citet{meu99} law: $A_{1600}= 4.43 + 1.99 \beta$, and considered a conversion $A_{6563} \approx A_{1600}/3.2$. We applied this new extinction correction (which we considered to be the same for the continuum and line) to each galaxy, and we re-computed its $\rm H\alpha$ rest EW and luminosity, and derived the corresponding SFR.

Fig.~\ref{fig_meurext} is the analogue of Fig.~\ref{fig_twomet}, but in this case we show the SFR-$\rm M^\ast$ plane and sSFR distribution that result from considering the new dust extinction correction (at fixed solar metallicity, as in Section~\ref{sec:results}). The star-formation main sequence and starburst cloud are clearly distinct in these plots, showing that the sSFR bimodality is still present when we consider a single, $\beta$-dependent extinction relation rather than the best-fit extinction obtained from the individual SED fitting of each of our galaxies.  We conclude that the observed sSFR bimodality is robust to reasonable changes in the modelling assumptions.

%% This command is needed to show the entire author+affilation list when
%% the collaboration and author truncation commands are used.  It has to
%% go at the end of the manuscript.
%\allauthors

%% Include this line if you are using the \added, \replaced, \deleted
%% commands to see a summary list of all changes at the end of the article.
%\listofchanges

\end{document}